# Equation-of-State, Critical Constants, and Thermodynamic Properties of Lithium at High Energy Density

Mofreh R. Zaghloul
Department of Physics, College of Sciences, United Arab Emirates University,
P.O.B. 15551, Al-Ain, UAE.

Ahmed Hassanein
Center for Materials under Extreme Environment (CMUXE),
College of Nuclear Engineering, Purdue University
500 Central Drive, West Lafayette, IN 47907-2017, USA

**ABSTRACT**

High-density lithium plasmas are expected to be generated in the Inertial Confinement Fusion (ICF) reactor chamber due to energy deposition of the prompt X-rays and ion debris in the first wall. These dense plasmas are encountered in many other applications as well. The design and optimization of such lithium-based applications requires information about the thermodynamic properties of Li fluid over a wide range of parameters including the high-energy-density regime.

A model that takes into account all essential features of strongly-coupled plasmas of metal vapors is developed and used to predict the equation-of-state (EOS), critical constants, thermodynamic functions, and principal shock Hugoniot of hot dense lithium fluid. Predictions and computational results from the present model are analyzed and presented and are available in electronic form suitable for use to interested researchers and scientists.

# 1. INTRODUCTION

Lithium is an alkali metal of substantial scientific importance. It is proposed as the liquid wall layer and tritium breeding material in a number of Inertial Confinement Fusion (ICF) reactor-chamber designs [1,2]. The energy deposition of the thermonuclear target explosion products (prompt X-rays and ion debris) within a shallow depth, of the ICF first-wall lithium-layer is anticipated to ablate, heat, and ionize the material generating lithium plasma over a wide range of temperatures and densities including the high energy density regime. High-density lithium plasmas are produced in many other scientific and technological applications as well [3,4]. Reliable prediction of the thermodynamic properties of Li fluid, essentially in the challenging ultrahigh pressure regime, is needed for design trade-offs and optimization of such lithium-based devices and applications. Of particular interest is the prediction of the critical constants (critical temperature $T_C$, density $\rho_C$, and pressure $P_C$) of lithium. These properties are required for understanding the fundamental mechanisms for many relevant physical phenomena and processes such as explosive boiling which have been found to be the most relevant thermal response mechanism to the high heating rate from the X-rays photon energy deposition [5]. However, due to their high values, experimental measurements of the critical constants of lithium are not available up to now [6]. Yet, several theoretical estimates of Li critical parameters have been introduced by various authors in the literature. These estimates are based on different empirical bases such as the relation to the boiling point, to the vaporization enthalpy, or to the surface tension etc. A typical sample of these estimates is shown in Table 1.

**Table 1** A representative collection of predicted values of the critical properties of lithium in the literature.

| Reference | $T_C$ [K] | $P_C$ [MPa] | $\rho_C$ [kg m$^{-3}$] |
|---|---|---|---|
| Dillon et al. 1966 [7] | 3223±600K | -- | 120±33 |
| Ohse et a. 1985 [8] | 3503±10 | 38.42±0.54 | 110.4±0.5 |
|  | 3344±42 | 30.4±2.0 | 118±2 |
| Fortov et al. 1990 [9] | 3223 | 69.80 | 105 |
| Hess 92 [6] | 4176 | 84.0 | -- |
| Gathers 1994 [10] | 3741.2 | 114.28 | 90.917 |
| Likalter et al. 2002 [11] | 3350 | 43.56 | 53 |

A plasma system may be theoretically described using a "physical" or a "chemical" model of the substance ("physical picture" or "chemical picture"). In the rigorous physical model, only free electrons and bare nuclei are considered with pure Coulomb inter-particle interaction. Clusters of elementary particles can be described as manifestations and consequences of nonideality or interaction between electrons and nuclei. On the other hand, most practical calculations of the equation-of-state (EOS) and thermodynamic properties for real substances are performed within the framework of the chemical model or "chemical picture". In the chemical model, all structured composites that have internal degrees of freedom are considered as elementary constituents of the ensemble together with free electrons and bare nuclei. The system can subsequently be fully described by specifying its thermodynamic potential. At constant density and temperature, the Helmholtz free-energy function is used as the thermodynamic potential and under equilibrium conditions, this thermodynamic potential is minimized.

With a statistical-mechanically-consistent evaluation of internal partition functions of various species, appropriate consideration of strong Coulomb-coupling among charged particles, and careful implementation of partial degeneracy and strong hard-sphere repulsion we extend the applicability of the "chemical picture" of plasma systems to the strong nonideal regime and use it to predict the properties of lithium at high energy density. Special attention is given to the phenomena of pressure dissociation and pressure ionization in the calculation of equilibrium composition of lithium species.

Establishing confidence and reliability in theoretically generated EOS and thermodynamic properties necessitates validation through comparison with available experimental measurements and other available independent theoretical predictions.

## 2. THEORETICAL THERMODYNAMIC MODEL

The appropriate description of a closed statistical ensemble is given by the canonical partition function or by its associated thermodynamic potential, the Helmholtz free energy function, $F$. The many-body partition function is defined in terms of the Hamiltonian of the system. Assuming that the momentum and potential functions of the Hamiltonian are uncoupled, the many-body partition function can be factorized into translational, internal, and configurational factors. The associated thermodynamic potential function, $F$, is therefore additive and can be regarded as the sum of translational, internal, and configurational components. With the additivity assumption of the thermodynamic potential and the aid of a physically-justified, statistical-mechanically-consistent scheme for smooth truncation of the internal partition functions, the chemical model allows the expression of bulk-state properties of the assembly in terms of relevant properties of individual particles where interactions or couplings among different species are usually accounted for in the configurational free energy component.

For the range of high temperatures considered herein, from slightly below $T_C$ to very high temperatures, the formation and existence of charged clusters can be approximately ignored and the lithium vapor may be effectively regarded as a mixture of lithium dimers, atoms, and atomic ions ($Li_2$, $Li_0$, $Li_{+1}$, $Li_{+2}$) at different excitation states, in addition to bare nuclei ($Li_{+3}$) and free electrons ($e$).

The Helmholtz free energy function, $F$, for this assembly can be written as;

$$F - F_{0K} = F_{Li2} + \sum_{s=0,\ldots,+3} F_{s,ideal} + F_e + F_{rad} + F_{non\_id} + F_{zero-of-energy} \qquad (1),$$

where $F_{0K}$ is the zero-Kelvin free energy, $F_{Li2}$ is the free energy of $Li$ dimers, $F_{s,ideal}$ is the classical ideal free energy of atomic particles of charge multiplicity $s$, $F_e$ is the free energy of free electrons, and $F_{non\_id}$ is the nonideal component of the free energy function which takes into account nonideal effects including Coulomb-coupling, van der Waals' corrections, strong hard-sphere repulsion and the exclusion of the occupied volume from that accessible to bare nuclei and free electrons. The term $F_{rad}$ in Eq. (1) is the free energy of the photon gas which is essential for very high temperatures (>$10^6$ K) while the $F_{zero-of-energy}$ term amends for the fact that all the free energy components must be referred to the same reference or "*zero-of-energy*".

For systems of temperatures considerably higher than the critical temperature, the zero-Kelvin free energy term is negligible compared to the thermal contribution on the right hand side of the equation. However, for temperatures in the vicinity of or below the critical temperature, this term is important and should be taken into account.

The equilibrium state composition of the system is determined through minimizing the free energy function, $F$ with respect to the occupational numbers $\{N_j\}$ subject to electroneutrality and conservation of nuclei. This minimization can be achieved iteratively through using optimization algorithms [12-14] or by casting the set of minimization equations, analytically, into the form of nonideal Saha-like equations which can then be solved using techniques and algorithms developed for this purpose [15,16]. A brief discussion of the advantages and disadvantages of both techniques for the minimization of the function can be found in [17].

Once the equilibrium composition is determined, the pressure $P$ and internal energy $U$ can be found from the relations

$$P = -\left(\frac{\partial F}{\partial V}\right)_{\{N\},T} \tag{2},$$

and

$$U = -T^2 \left(\frac{\partial F/T}{\partial T}\right)_{\{N\},V} \tag{3},$$

All other thermodynamic properties can be determined from $P$ and $U$ using standard thermodynamic relations. It should be noted that one may use the free energy formulation of the vapor phase to represent the liquid phase of the material in the neighborhood of the critical point where atoms at such high temperatures tend to move freely through the body in a way similar to their random motion in the gas phase.

## 2.1. Zero-Kelvin Isotherm

Zero-Kelvin (0K) isotherms are needed to develop a wide-range EOS model for lithium or any other material. Recently, Young et. al. [18] used simple lattice dynamics models to reduce the compiled experimental measurements of compression isotherms of all naturally occurring elements to the 0K. For lithium the compression data were taken from Guillamue et al. [19]. The generated data are up to 100 GPa compression. As noted by the authors, the compilation is not error free; however, this compilation is the largest available to us to use in this study. It has to be noted that there are several distinct solid phases of lithium below 100 GPa and a great deal of effort has been devoted into identifying them in recent years. In fact, some uncertainty has recently been expressed about the proper ambient phase of lithium (see Ackland, et al. [20]). A thorough model would take these details into account, however, as the present model is concerned with lithium fluid from the neighborhood of the critical point to very high temperatures, simple fitting to the compilation of Young et al. seems to be sufficient or at least a reasonable starting point for this work.

In contrast to the case of Li-compression, information about the rarefaction region of lithium at 0K is limited. Shultz [21] measured the yield strength for lithium using the 0.2% offset rule to be 81 psi (0.5585 MPa). As indicated by Shultz, the value for the ultimate strength or "tensile strength" for lithium was listed in [22], as an inequality, to be <2200 psi (<15.2 MPa). This value is relatively small when compared to the local thermal pressure for the range of temperatures considered in this study (from slightly below the critical point to very high temperatures) and one may marginally consider ignoring the negative part of the "cold curve" pressure for lithium fluid at these temperatures and for densities less than the normal density. However, the sudden change from the compression curve to zero pressure for densities less than the normal density at 0K may cause or produce undesirable discontinuities in the thermodynamic properties. A better choice, therefore, is to represent the compression data by an analytical expression that produces small negative pressure (comparable to the tensile strength) for the rarefaction region.

From a number of empirical fitting formulae available in the literature we considered the following relatively simple ones:

(*i*) Boettger et. al. [23] fitted the fcc results calculated using the Kohn-Sham-Gaspar (KSG) and Rajagopal-Singhal-Kimball (RSK) potentials to the Murnaghan equation,

$$P_{0K} = (B_0/B_0') \left( \left(\frac{V_0}{V}\right)^{B_0'} - 1 \right) \tag{4},$$

where $V_0$ and $B_0$ are the specific volume and bulk modulus while $B_0' = dB/dP$, all at the reference condition ($P=0$ and $T=0$). The following values of the parameters for lithium are given: KSG ($V_0$=137.8058 a.u or 0.001772 m$^3$/kg, $B_0$=170.90978 kbar, and $B_0'$=2.39004) and RSK ($V_0$=122.6615 a.u or 0.0015773 m$^3$/kg, $B_0$=208.99432 kbar, and $B_0'$=2.43219);

(*ii*) Khishchenko [24] represented the Helmholtz free energy for the 0K lattice by the expression,

$$F_{0K} = \frac{B_0 V_0}{m-n} \left( \frac{1}{m}\left(\frac{V_0}{V}\right)^m - \frac{1}{n}\left(\frac{V_0}{V}\right)^n \right) + E_{sub} \tag{5},$$

where the quantity $E_{sub}$ is the sublimation energy and is determined from the condition $E_{0K}(V_0)=0$ to be

$$E_{sub} = \frac{B_0 V_0}{m\,n} \tag{6},$$

where $E_{0K}$ is the 0K isotherm of the internal energy. For lithium, the parameters in the above equations are given as $m$=0.67, $n$=0.48, $V_0$=0.0018422 m$^3$/kg, and $B_0$=11.887 GPa. The internal energy and the pressure isotherm at 0K are given by the general thermodynamic relations in (2) and $E = F$ at 0K. The pressure is given, therefore, by

$$P_{0K} = -\left(\frac{\partial F_{0K}}{\partial V}\right)_T = \frac{B_0 V_0}{m-n}\left(\left(\frac{V_0}{V}\right)^m - \left(\frac{V_0}{V}\right)^n\right) \tag{7};$$

(*iii*) In developing a high temperature equation-of-state of aluminum, Naumann [25] proposed the following form for the Helmholtz free energy for the 0K lattice;

$$F_{0K} = L_{v0}\left(e^{2b(1-\xi)} - 2e^{b(1-\xi)}\right) \tag{8},$$

where $L_{v0}$ is the 0K heat of vaporization or total binding energy, $\xi = \left(\frac{V}{V_0}\right)^{1/3}$ and $b$ is an empirical constant. Again, the 0K isotherm of the internal energy takes the same form as $F_{0K}$ while the 0K pressure isotherm is given by

$$P_{0K} = -\left(\frac{\partial F_{0K}}{\partial V}\right)_T = \frac{2L_{v0}b}{3V_0\xi^2}\left(e^{2b(1-\xi)} - e^{b(1-\xi)}\right) \quad (9),$$

This formula can be used to represent the data compiled by Young et. al.[18] with appropriate choice of the parameter $b$;

(*iv*) Another useful formula is the Vinet exponential formula [26] where the 0K pressure is expressed as

$$P_{0K} = 3B_0\left[1 - \left(\frac{V}{V_0}\right)^{1/3}\right]\left(\frac{V}{V_0}\right)^{-2/3} exp\left\{\frac{3}{2}(B_0' - 1)\left[1 - \left(\frac{V}{V_0}\right)^{1/3}\right]\right\} \quad (10),$$

The corresponding 0K internal energy (or the Helmholtz free energy) curve can therefore be written in the form

$$F_{0K} = E_{0K} = E_0 + \frac{4B_0V_0}{(B_0'-1)^2}\left\{1 - \left(1 + \frac{3}{2}B_0'\left[\left(\frac{V}{V_0}\right)^{\frac{1}{3}} - 1\right] - \frac{3}{2}\left(\frac{V}{V_0}\right)^{\frac{1}{3}}\right)exp\left(\frac{3}{2}((B_0' - 1) \times \left[1 - \left(\frac{V}{V_0}\right)^{1/3}\right]\right)\right\} \quad (11),$$

where at the normal volume, $E_{0K}(V_0) = E_0 + 6B_0V_0/(B_0' - 1)^2$.

These empirical fitting formulae have been plotted against the compilation of Young et al. [18] as shown in Figs. 1-a and 1-b. As it can be seen from Fig. 1-a, all formulae may be considered reasonable representation of the compiled data. The slight spread between the formulae and the compilation of Young et al. could be due to the fact that none of the theories behind these formulae considered the many different solid phases of lithium. On the other hand, all of the above formulae have the common problem of overestimating the tensile strength of lithium (rarefaction region) with Vinet exponential formula behaving best in this regard among all formulations. Accordingly, in the present model, we adopt Vinet's exponential formula for the 0K isotherms with the following values for the parameters $B_0$=6.0 GPa and $B_0' = 5.0$. The corresponding internal energy curve will depend on the choice of the energy-zero point or in other words on $E_0$. Choosing $E_0 = -6B_0V_0/(B_0' - 1)^2$ means that the standard state is the lithium molecules in the solid phase at the normal volume and 0 K. On the other hand, choosing $E_0$ equal to the negative of the sum of phase change energies, dissociation energy and $6B_0V_0/(B_0' - 1)^2$ means that the chosen standard state is the state of dissociated and separated atoms in the vapor phase at 0K. In this work we choose the molecules in the solid phase at the normal volume and 0K as the standard state.

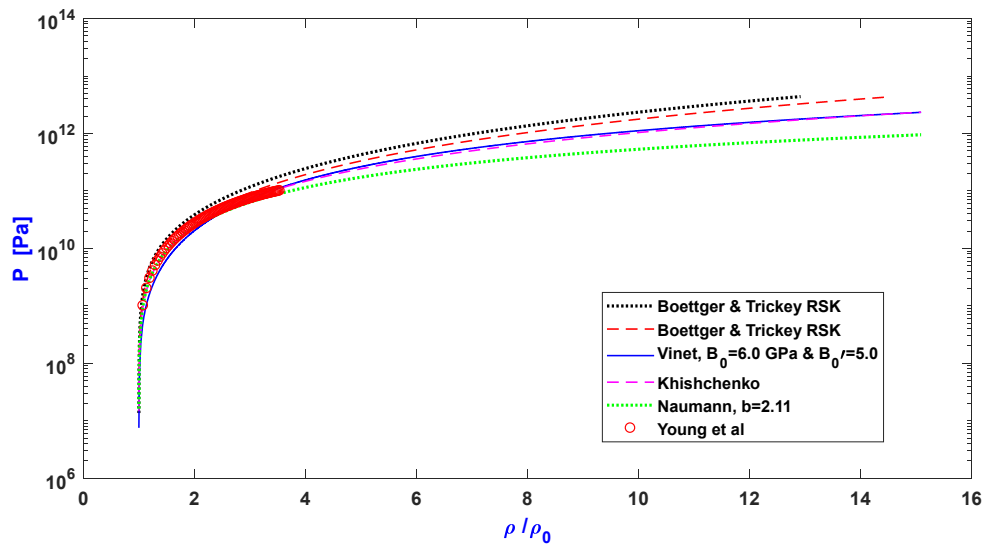

*a*

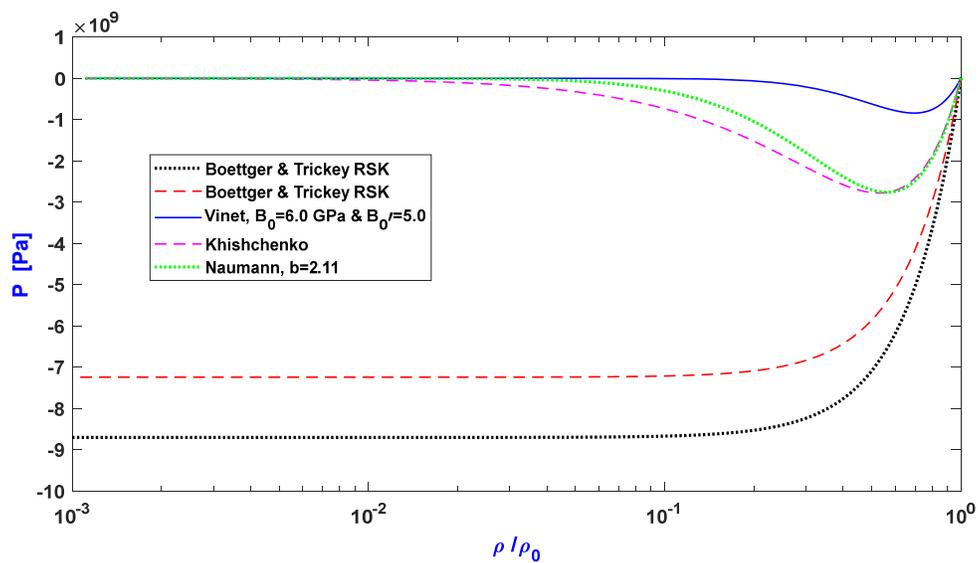

*b*

Fig. 1: (a) Comparison between the compilation of Young et al. for the compression data of lithium at 0K and a set of representative formulae, (b) Behavior of 0K formulae in the negative pressure region.

## 2.2. Free Energy Components

Because of their heavy masses, heavy particles remain classical even at very high density and the Helmholtz free energy function for non-interacting lithium dimers can be written as

$$F_{Li2} = -K_B T N_{Li2} \left( ln\left[\frac{Q_{total,Li2}}{N_{Li2}}\right] + 1 \right) \tag{12},$$

where the total partition function for the dimers, $Q_{total,Li2}$, can be factored as

$$Q_{total,Li2} = \left(\frac{2\pi m_{Li2} K_B T}{h^2}\right)^{\frac{3}{2}} V \, Q_{vib} \, Q_{rot} \, Q_{elec,Li2} \tag{13},$$

with the first factor in the right hand side of the above equation, $\left(\frac{2\pi m_{Li2} K_B T}{h^2}\right)^{\frac{3}{2}} V$, representing the translational partition function while $Q_{vib}, Q_{rot}$ and $Q_{elec,Li2}$ are the vibrational, rotational, and electronic factors, respectively. Only at temperatures sufficiently lower than the dissociation energy of the molecule, lithium dimers can exist in appreciable amounts. Hence, retaining the ground electronic state of the molecule may be considered as a reasonable approximation. The vibrational-rotational factor, $Q_{vib} Q_{rot}$, can be evaluated as in Ref. [27] using the spectroscopic data taken from Huber and Herzberg [28].

The free energy of atomic particles of charge multiplicity $s$ is given by a similar classical expression,

$$F_{s,ideal} = -K_B T N_s \left( ln\left[\left(\frac{2\pi m_s K_B T}{h^2}\right)^{\frac{3}{2}} \frac{V Q_{int,s}}{N_s}\right] + 1 \right), \qquad s = 0,1,2,3 \tag{14},$$

where $T$ is the absolute equilibrium temperature, $K_B$ is the Boltzmann constant, $m_s$ is the mass of the particle *s*, *h* is Planck's constant, *V* is the volume of the system and $Q_{int,s}$ is the sum over states (internal partition function) for the particle *s*. For bare nuclei $Q_{int,3} = 1$.

For an assembly of partially degenerate electrons, the free energy of an ideal Fermi gas can be used where,

$$F_e = -P_e V^* + N_e \mu_{e,id}^*$$

$$= -\left(\frac{2K_B T V^*}{\Lambda_e^3}\right) I_{\frac{3}{2}}\left(\frac{\mu_{e,id}^*}{K_B T}\right) + N_e \mu_{e,id}^* \qquad (15).$$

In Eq. (15), $V^*$ is the volume accessible to free electrons after excluding the portion occupied by extended particles, $\Lambda_e$ is the average thermal wavelength of electrons, $P_e$ is the electron pressure and $\mu_{e,id}^*$ is the chemical potential of the ideal Fermi electron gas, while $I_\nu$ is the complete Fermi-Dirac integral defined as,

$$I_\nu(\vartheta) = \frac{1}{\Gamma(\nu+1)} \int_0^\infty \frac{y^\nu}{\exp(y-\vartheta)+1} \, dy \qquad (16).$$

The electron chemical potential $\mu_{e,id}^*$ is related to the number of free electrons in the system by

$$N_e = (2V^*/\Lambda_e^3) I_{\frac{1}{2}}\left(\frac{\mu_{e,id}^*}{K_B T}\right) \qquad (17).$$

The evaluation of the internal partition function for the composite structured particles has been thoroughly discussed and investigated in Refs [17,29-32]. An essential concern in modeling thermodynamic equilibrium at high density is associated with the need to consistently and smoothly truncate the internal partition functions of the ensemble of bound states. The obsolete remedy of this problem by an abrupt *step-cutoff* of the excited states with energies higher than the perturbed (lowered) ionization energy is known to produce discontinuities and singularities in the derivatives of the free energy and in the thermodynamic functions despite the smooth variation in the external thermodynamic parameters (density and temperature). Instead, *continuous* state-dependent weights or occupational probabilities, $w_i(T,V,\{N\})$, usually derived from the plasma microfields, and which decrease monotonically and indefinitely with density are used for smooth and continuous transition from bound to free electronic states.

One possible problem with the evaluation of the internal partition functions, when using continuous occupational probabilities that decrease monotonically and indefinitely with density, is the possible cutoff of the ground states at extreme high densities causing vanishing of the internal partition function with the collapse of the foundation of the whole statistical problem in this case. To avoid this problem, the conventional electronic partition function with energy states relative to the electronic ground, $Q_{int}(T,V,\{N\})$, is determined using the expression proposed in [32];

$$Q_{int}(T,V,\{N\}) = [1 - w_0(T,V,\{N\})] + \sum_{i=0}^{\infty} g_i w_i(T,V,\{N\}) \, e^{-\varepsilon_i/K_B T} \qquad (18),$$

where $g_i$ is the statistical weight of the *i-th* level whose energy relative to the ground is $\varepsilon_i$, and $w_i(T,V,\{N\})$ is the state-dependent occupational probability of the *i-th* level. As it can be seen, the from (18) retains at least one state of the ground level in a strongly perturbed system where $w_0(T,V,\{N\})$ goes to zero at extreme high densities and low temperatures. The simple expression given by Salzmann [33] for the occupational probabilities is adopted in the present model.

The contribution of the photon gas to the thermodynamics of the system becomes significant at relatively high temperatures. For a blackbody radiation, the free energy is expressed as [34]

$$F_{rad} = -(4\sigma/3c)VT^4 \qquad (19),$$

where $\sigma$ and $c$ are the Stefan-Boltzmann constant and the speed of light, respectively.

## 2.3. NONIDEAL AND ZERO-OF-ENERGY COMPONENTS

The nonideal free energy term takes into account the Coulomb coupling among charged particles, $F_{Coul}$, van der Waals' correction, $F_{vdw}$, strong hard sphere repulsion, $F_{hs}$, in addition to the exclusion of the occupied volume from that accessible to bare nuclei, $F_{bn}$, that is

$$F_{conf} = F_{Coul} + F_{vdw} + F_{hs} + F_{bn} \qquad (20).$$

The long-range Coulomb interactions can be taken into account through an approximate ion-sphere or cellular model in which the system is regarded as a number of electrically neutral spherical unit cells. Ions located at the centers of the cells neutralize the background electrons inside the unit cell. A Coulomb correction term to the free energy can therefore be written as

$$F_{Coul} \approx -\frac{9}{10} \frac{e^2}{4\pi\varepsilon_0 R_0} \sum_s N_s s^2 \qquad (21),$$

where $R_0$ is the cell radius.

The van der Waals' term of the configurational free energy takes into account the attractive part of the potential between neutral particles which becomes appreciable at low temperatures. A very simplified treatment of this term can be expressed as

$$F_{vdw} \approx -\frac{\pi}{3} \frac{N_0^2}{V} (\delta_0 \sigma_0^3) \qquad (22),$$

where $\delta_0$ is the van der Waals' well depth and $\sigma_0$ is the distance at which the interaction energy is zero.

The part of the nonideal free energy that takes into consideration the effect of the finite size of extended particles is usually expressed in terms of the packing fraction, $\eta$. Expressions for this term can be found elsewhere [35,36].

The final term of the nonideal energy takes into account the reduction in the accessible volume for bare nuclei and can be expressed as

$$F_{bn} = -K_B T N_3 \ln(1-\eta) \quad (23).$$

The calculation of the packing parameter, $\eta$, requires information about the radii of all finite-size particles employed in the model. It is impractical to consider a few thousand of species in different excitation states of different radii in the calculation of the packaging parameter and hard-sphere component of the configurational free energy. Instead, an average hard-sphere radius is assigned and used for each ion characterized by a particular charge state. The hard-sphere radii, $r_s$, for ions of charge state, $s$, are generally unknown, particularly for the higher charged ions. However, it is assumed that ionic radii decrease monotonically with increasing ion multiplicity. Accordingly, one can associate the radii of ions of different multiplicities to a certain reference radius like the atomic radius, $r_0$. In the present study, we use for the radius of ions of multiplicity $s$ the approximate formula [37];

$$r_s = \left(\frac{Z_{nuc}-s}{Z_{nuc}}\right)^n r_0 \quad (24),$$

where $n$ is a parameter to be investigated.

Different estimates for vacuum or low density values of $r_0$ exist in the literature. The choice from among these estimates is left to agreement with available experimental measurements. At high densities, however, consistency with the cellular model used for Coulomb interactions prescribes ionic radius smaller than the Wigner-Seitz (density dependent) radius. Abiding by this consistency requirement, a smoothing formula is used to smoothen the transition between the low density constant value $r_{0,0}$ and the density-dependent Wigner-Seitz radius where

$$r_0(\rho) = \left(r_{WS}^{-3} + r_{0,0}^{-3}\right)^{-1/3} \quad (25).$$

Keeping in mind that the standard state chosen for these calculations is the state of molecules in the solid phase at $V_0$ and 0K, the $F_{zero-of-energy}$ term will, therefore, embody the phase change, dissociation, and ionization energies, i.e.,

$$F_{zero-of-energy} = L_F + L_V + \sum_{r=0}^{3} N_r (D_0/2) + \sum_{s=1}^{3} N_s \sum_{\zeta=1}^{s} I_{\zeta-1} \qquad (26),$$

where $L_F$ is the latent heat of fusion, $L_V$ is the latent heat of vaporization, $D_0$ is the dissociation energy while $I_\zeta$ is the ionization energy for the ion $\zeta - 1$.

## 2.4. ATOMIC RADIUS

The fact that the boundary of the electron cloud surrounding the atomic nucleus is not a well-defined physical quantity led to various definitions of the atomic radius (van der Waals, covalent, metallic, etc.). Van der Waals radius is principally defined as half the minimum inter-nuclear separation of two non-bonded atoms of the same element [38]. The covalent radius, on the other hand, is defined as the radius of the atoms of an element when covalently bound to other atoms. It can be deduced from the separation between the atomic nuclei in molecules as the length of that covalent bond is equal to the sum of their covalent radii. Similarly, the metallic radius is defined as the radius of the atom when connected to other atoms by metallic bonds. Other definitions also exist and the value of the atomic radius generally depends on the atom's state and context. In the present calculations the metallic radius of lithium, calculated from the density, is used where $r_{0,0} \approx 1.53$ Å.

## 2.5. SOLUTION ALGORITHM

Among different chemical species in the ensemble there exist a set of chemical reactions including dissociation, ionization, and the inverse recombination processes. At equilibrium, the free energy function (Eq. (1)) is minimized for each of the possible reactions. For example, minimizing the free energy for the dissociation process $Li_2 \leftrightarrow 2Li_0$ requires

$$\frac{\partial F}{\partial N_{Li2}} - 2\frac{\partial F}{\partial N_{Li0}} = 0 \qquad (27).$$

Similarly, the minimization of $F$ for the ionization equilibrium reaction $Li_\zeta \leftrightarrow Li_{\zeta+1} + e$ requires that

$$\frac{\partial F}{\partial N_{Li,\zeta}} - \frac{\partial F}{\partial N_{Li,\zeta+1}} - \frac{\partial F}{\partial N_e} = 0 \qquad (28).$$

In Eqs. (27, 28) $N_{Li2}$, $N_{Li0}$, $N_{Li,\zeta}$, and $N_e$ stand for the occupation numbers of lithium molecules, atomic lithium, lithium ions of multiplicity $\zeta$ and free electrons, respectively.

At very low temperatures, one can roughly assume that only lithium molecules exist. As the temperature increases a fraction $f$, of the original number of molecules $N_0$, is dissociated into atomic lithium. In general, this fraction $f$ is a function of both of the temperature $T$ and density $\rho$. At any time, the number of un-dissociated molecules in the mixture is $N_{Li2}=(1-f)N_0$ while the number of neutral lithium atoms becomes $N_{Li,0} = 2fN_0\alpha_0$ where $\alpha_0$ is the proportion of the neutral atoms, calculated from the ionization equilibrium, relative to the total number of atomic species (atoms plus ions) i.e., $\alpha_0 = N_{Li,0}/(N_{Li0}+N_{Li,+1}+N_{Li,+2}+N_{Li,+3})$.

The algorithm used to solve the problem is summarized in the flowchart shown in Fig. 2. Details about iterative solution of the chemical equilibrium problem and the ionization equilibrium problem are disregarded here to simplify the readability of the flowchart.

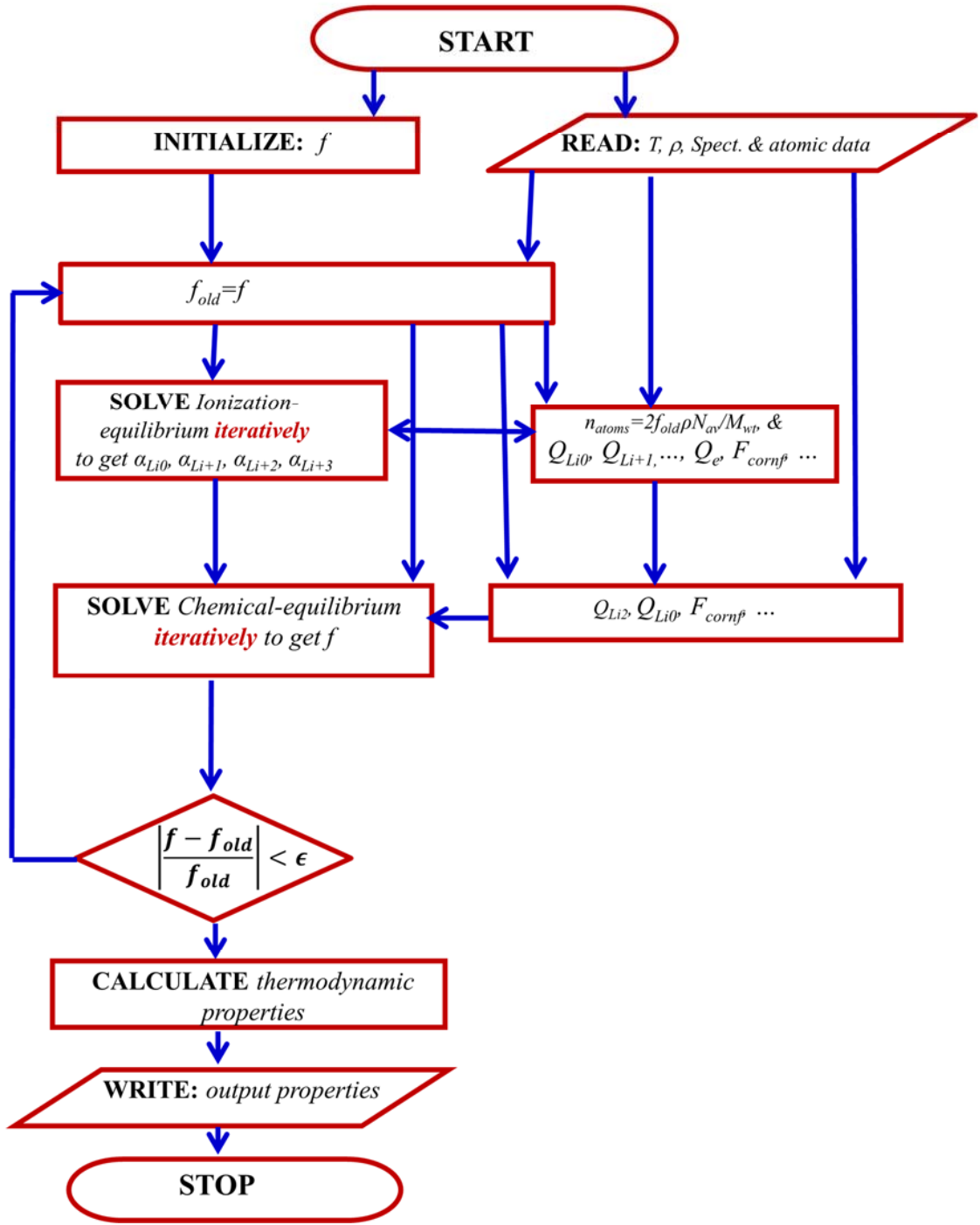

Fig. 2: Flowchart representing the computational algorithm used.

## 3. RESULTS AND DISCUSSIONS

Figure 3 shows a set of isotherms of the lithium dissociation fraction as a function of density. The dissociation is complete at low densities and/or high temperatures, as expected. However, it decreases in the region of intermediate densities and low temperatures due to the enhancement of recombination with density and the lack of sufficient energy for complete dissociation. At such low temperatures, further increase in density leads to enhancement in the dissociation due to high density nonideal effects (pressure dissociation).

Similar enhancement in ionization at intermediate to moderately high densities, due to Coulomb interactions among charged particles, is well reported and explained in the literature and can be clearly seen in Fig. 4, which shows a surface plot of the average degree of ionization, $Z_{av}$, of lithium as a function of density and temperature. The expected monotonic increase of the isochores of $Z_{av}$ with the temperature is clear from the plot as well. A plateau at $Z_{av}=1$ can be recognized due to the helium-like stable configuration.

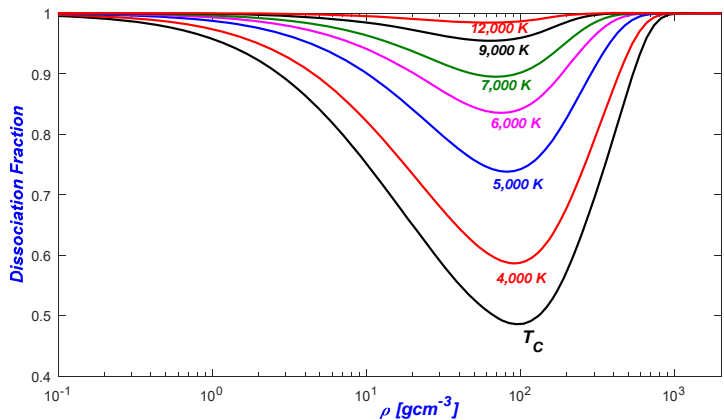

Fig. 3. Isotherms of the dissociation fraction of Li as a function of density.

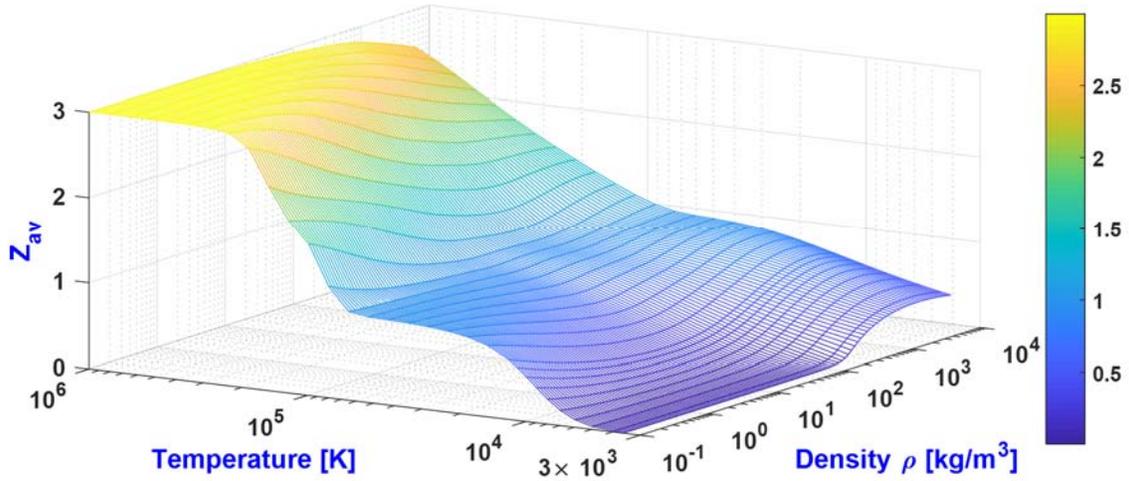

Fig. 4. A 3-D surface plot of the average ionization state of Li as a function of density and temperature.

At low temperatures (around the critical point), computational experiments of the present model showed that the model is insensitive to the parameter *n*, to which the radius of an ion depends on the number of remaining bound electrons though it is sensitive in this temperature domain, to the other two parameters, namely the radius of the neutral atom in vacuum $r_{0,0}$ and the parameter $(\delta_0 \sigma_0^3)$ in the van der Waals correction. Conversely, at high temperatures, model predictions are sensitive to *n* but not to $r_{0,0}$ and $(\delta_0 \sigma_0^3)$ as expected. Accordingly, in our search for the critical point we arbitrarily used the value *n*=0.8 while the metallic value of 1.53 Å is used for $r_{0,0}$ and the value for $(\delta_0 \sigma_0^3)$ is taken from [39]. These values of $r_{0,0}$ and $(\delta_0 \sigma_0^3)$, used in the present computations, are chosen based on consistency with available values of the density of the liquid-phase at temperatures below the critical point.

The liquid-vapor critical point is a point of inflection characterized by,

$(\partial P/\partial v)_T = (\partial^2 P/\partial^2 V)_T = 0$ (29).

This means that the isothermal compressibility is infinite at the critical point with the result that many other relevant properties of the substance would show singularity in the critical region. Identifying the critical point (which requires satisfying the conditions in (29)), can therefore be achieved through numeric inspection of the EOS isotherms.

It is worth mentioning that most chemical models of the EOS commonly show instabilities and develop loops in the liquid-vapor equilibrium region. The Maxwell equal-area construction is applied to correct for such anomalies for isotherms lower than the critical isotherm as shown in Fig. 5. The figure shows pressure isotherms of Li fluid as a function of density. Numerical search predicted the critical point to be characterized by ($T_C$ = 3495 K, $P_C$ = 0. 623 GPa, $\rho_C$ = 223.5 kg/m$^3$) (new values $T_C$ = 3084 K, $P_C$ = 0. 527 GPa, $\rho_C$ = 245.3 kg/m$^3$ ) for a value for the parameter $n=0.8$. It has to be noted that the value of $T_C$ predicted from the present work lies within the range of values in Table 1. However, the critical density predicted from the present study, $\rho_C$ = 223.5 kg/m$^3$ is higher than the values in Table 1 though considerably less than the density of liquid lithium near the normal boiling point ($\rho_b$ = 399.15 kg/m$^3$ at $T_b$ =1620 K [40]), as expected. In addition, although the present model is proposed for temperatures from the neighborhood of the critical temperature to very high temperatures, liquid densities calculated at 2500 K (353.1 kg/m$^3$) and 3000 K (327.7 kg/m$^3$) meet the expectations of getting a thermodynamic dome for the two phase region. The critical pressure predicted from the present study is considerably higher than those in Table 1, though still lower than he corresponding pressure for an ideal gas. However, the fact that the predictions in Table 1 are empirical in nature makes it difficult to derive solid quantitative conclusions about the absolute values of the critical parameters predicted from the present model. Even though, the values predicted herein are undoubtedly consistent with the associated predictions of the thermodynamic properties.

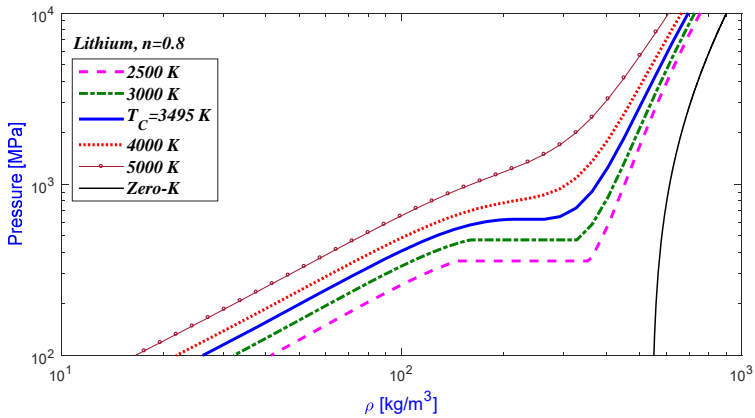

Fig. 5: Pressure Isotherms around the critical point of lithium.

Figure 6 shows a larger set of isotherms including the zero-K and critical isotherms followed by isotherms from 4000 to 12000 K with an increment of 1000 K, then by isotherms from 15000 to $10^9$ K which are equally spaced on the logarithmic scale with $\Delta log_{10}(T)=0.160768$. These isotherms are generated neglecting the contribution of the photon gas. As expected, the pressure isotherms of the vapor phase behave ideally at low densities and/or high temperatures. However, deviations from this ideal behavior develop for high densities and relatively low temperatures and become most prominent with the critical isotherm.

Figure 7 presents the corresponding isotherms for the specific internal energy of the lithium fluid where similar observations can be noted.

At the critical temperature, $T_C$, many physical properties of pure fluids exhibit special behavior near the critical point. For example, the ratio of specific heats, γ, of fluid becomes infinitely large at the critical point as it appears clearly in Fig. 8 where γ increases sharply as the temperature gets closer to the critical temperature.

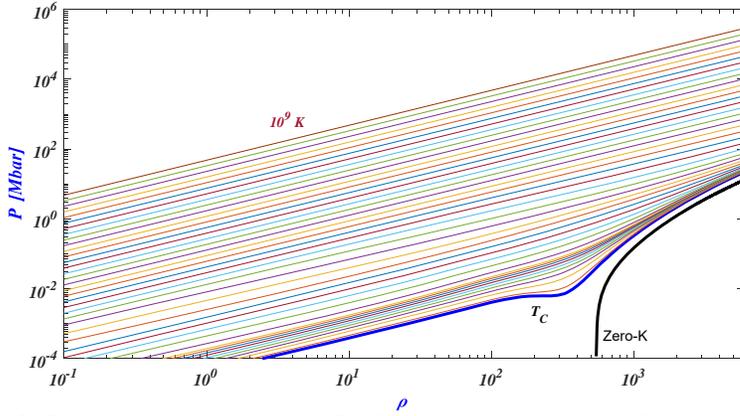

Fig. 6: Pressure isotherms of Li fluid as a function of density; the lowest isotherm is the zero-K isotherm followed by the critical isotherm ($T_C$ = 3566 K) for $n=0.8$, then isotherms from 4000 to 12000 K with an increment of 1000 K followed by isotherms from 15000 to $10^9$ K which are equally spaced on the logarithmic scale with $\Delta log_{10}(T)=0.160768$.

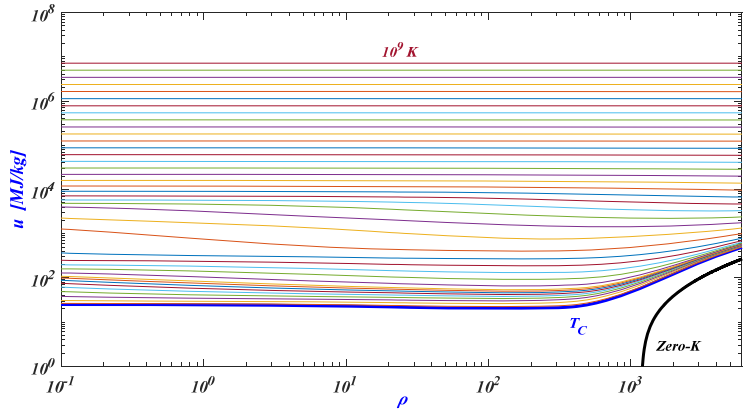

Fig. 7: Isotherms of the specific internal energy of Li fluid as a function of density (same set of isotherms as in Fig. 6).

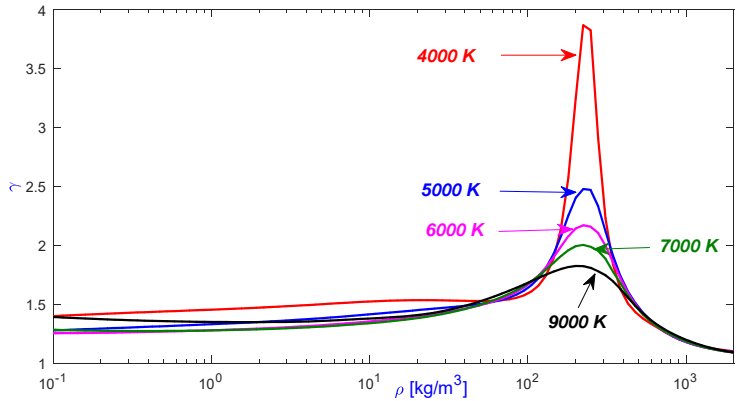

Fig. 8: Isotherms of the ratio of specific heats showing singularity near the critical point.

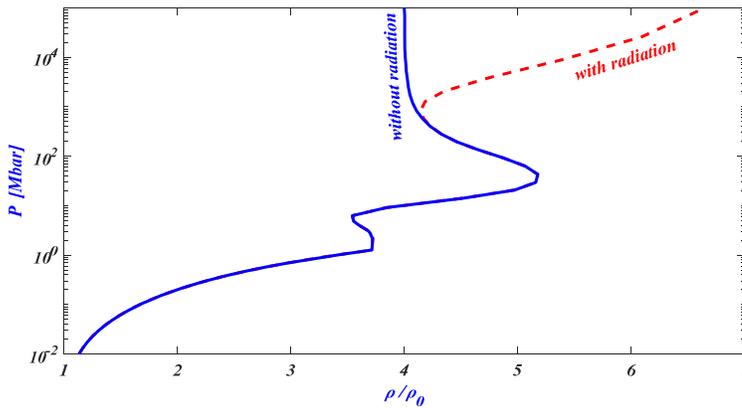

Fig. 9: Principal shock Hugoniot of Li calculated from the present equation-of-state with (solid line) and without radiation (dashed line).

Figure 9 shows the principal shock Hugoniot of Li as predicted from the present EOS (with and without radiation) together with the zero-K data. Ignoring the contribution of radiation, extremely high pressures and temperatures developed behind very strong shock waves fully ionize the lithium plasma, which approaches an ideal gas of bare nuclei and free electrons. For these conditions, the Hugoniot curve approaches its theoretically-expected asymptotic value for an ideal gas ($\rho/\rho_0=4$). For the region of intermediate to high pressures where the material is partially ionized and the electronic structure determines the equation-of-state, a maximum compression beyond the ideal gas asymptote is recognized and the lithium Hugoniot shows a kink or shoulder in this region. This kink appears as a result of the competition between the absorption of internal energy for exciting and ionizing the electrons in the *K* shell (enhances compression) and the increase in pressure due to release of more free electrons due to ionization, which opposes compression. For lower pressures (weaker shock), predictions from the present model for EOS of Li show a second lesser kink due to ionization of the *L*-shell electrons (one per atom) in the medium. As the *K*-shell in the Li atoms is filled, it generally requires high ionization energies than the *L*-shell, which has only one electron and a jump exits between the ionization potentials of the electron in the *L*-shell and the first electron of the *K*-shell. Because of this jump in the ionization potentials, two levels of shock strengths would be required for the removal of the electrons of each shell.

Considering the contribution of radiation, the shock Hugoniot of lithium, as shown by the dashed curve in Fig. 9, approaches the well-known theoretical limiting value of $\rho/\rho_0 = 7$ characteristic to the photon gas.

## 4. CONCLUSIONS

$T_C = 3084$ K, $P_C = 0.527$ GPa, $\rho_C = 245.$ kg/m$^3$

Estimates of the critical properties ($T_C = 3495$ K, $P_C = 0.623$ GPa, $\rho_C = 223.5$ kg/m$^3$) of pure lithium fluid are obtained from a chemical model for the equation-of-state (EOS) of hot dense partially ionized fluid. The model is used to generate a consistent set of thermodynamic properties of lithium fluid and to predict the principal shock Hugoniot curve. The calculated Hugoniot curves show the expected limiting behavior of an ideal gas

Page 26

$\rho/\rho_0)_{max}$ =4 and of a photon gas $\rho/\rho_0)_{max}$ =7 for very strong shocks (extremely high temperatures and high pressures). The principal Hugoniot calculated from the present model shows two shoulders or kinks due to ionization of the electrons in the *K* and *L* shells. The present prediction of the critical temperature of Li lies in the range of values reported in the literature, however, the predictions of the critical density and critical pressure are higher than those reported from empirical methods in the literature.

## ACKNOWLEGMENTS

This work is supported by the UAEU-UPAR Project, contract G00002907.

## REFERENCES

[1] W. R. Meier, Liquid Wall Chambers, LLNL-TR-471596, 2011.

[2] R. W. Moir, Liquid Wall Chambers for HIF, 19th International Symposium on Heavy Ion Inertial Fusion (HIF2012).

[3] W. Theobald, C. Wülker, J. Jasny, J. S. Bakos, J. Jethwa, F. P. Schäfer, "High-density lithium plasma columns generated by intense subpicosecond KrF laser pulses," Optics Communications Volume 149, Issues 4–6, 15, 289-295 (1998).

[4] William Whitty, John Costello, Eugene T. Kennedy, Christopher Moloney , and Jean-Paul Mosnier, "Absorption spectroscopy of an expanding laser produced lithium plasma in the extreme ultraviolet using the Dual Laser Plasma technique ," Applied Surface Science 127, 686-691 (1998).

[5] Mofreh R. Zaghloul and A. Renė Raffray, "IFE Liquid Wall Response to The Prompt X-Ray Energy Deposition: Investigation of Physical Processes and Assessment of Ablated Material," Fusion Science and Technology Vol. 47, 27-45 (2005).

[6] H. Hess, "Critical Point of Lithium under Influence of Coulomb Interaction," in <u>Physics of Nonideal Plasmas</u>, edited by W. Ebeling, A. Forster, and R. Radtk, Teubner Verlagsgesellschaft, Stuttgart (1992).

[7] I. G. Dillon, P. A. Nelson, and B. S. Swanson, "Measurement of Densities and Estimation of Critical Properties of the Alkali Metals," J. Chem. Phys. 44, 4229-4238 (1966).

[8] R. W. Ohse, J.-F. Babelot, J. Magill And M. Tetenbaum, "An Assessment of the Melting, Boiling, and Critical Point Data of the Alkali Metals," Pure & Appl. Chem., Vol. 57, No. 10, pp. 1407-1426 (1985).

[9] V. E. Fortov and I. T. Yakubov, <u>Physics of Nonideal Plasma</u>, Hemisphere, New York, (1990); [Russ. original] Chernogolovka (1984).

[10] G. Roger Gathers, <u>Selected Topics in Shock wave Physics and Equation-of-State Modeling</u>, World Scientific Publishing Co. Pte. Ltd. (1994).

[11] Alexander A Likalter, Helmut Hess and Hartmut Schneidenbach, "Critical Parameters of Alkali Metals: Deviation from Scaling," Physica Scripta, Volume 66, Number 1 (2002).

[12] G. M. Harris, J. E. Roberts, and J. G. Trulio, "Equilibrium Properties of a Partially Ionized Plasma," Physical Review, 119 (6), 1832-1841 (1960).

[13] D. G. Hummer and Dimitri Mihalas, "The Equation-of-State for Stellar Envelopes. I. An Occupation Probability Formalism for the Truncation of Internal Partition Functions," The Astrophysical Journal 331, 794-814 (1988).

[14] D. Mihalas, W. Däppen, and D. G. Hummer, "The Equation-of-State for Stellar Envelope. II Algorithm and Selected Results," The Astrophysical Journal, 331, 815-825 (1988).

[15] Mofreh R. Zaghloul, M. A. Bourham, and J. M. Doster, "A simple formulation and solution strategy of the Saha equation for ideal and nonideal plasmas," J. Phys. D: Appl. Phys., Vol. 33, 977-984 (2000).

[16]Mofreh R. Zaghloul, "Reduced Formulation and Efficient Algorithm for the Determination of Equilibrium Composition and Partition Functions of Ideal and Nonideal Complex Plasma Mixtures," Physical Rev. E., 69, 026702 (2004).

[17]M. R. Zaghloul, "On the Calculation of Equilibrium Thermodynamic Properties and the Establishment of Statistical-Thermodynamically-Consistent Finite Bound-State Partition Functions in Nonideal Multi-Component Plasma Mixtures within the Chemical Model," Phys. Plasmas, Vol.17, 122903-13 (2010).

[18]David A. Young, Hyunchae Cynn, Per Söderlind, and Alexander Landa, "Zero-Kelvin Compression Isotherms of the Elements $1 \leq Z \leq 92$ to 100 GPa," Journal of Physical and Chemical Reference Data 45, 043101 (2016).

[19]C. L. Guillaume, E. Gregoryanz, O. Degtyareva, M. I. McMahon, M. Hanfland, S. Evans, M. Guthrie, S. V. Sinogeikin, and H.-K. Mao, "Cold melting and solid structures of dense lithium," Nat. Phys. 7, 211 (2011).

[20]Graeme J. Ackland, Mihindra Dunuwille, Miguel Martinez-Canales, Ingo Loa1, Rong Zhang, Stanislav Sinogeikin, Weizhao Cai, Shanti Deemyad, "Quantum and isotope effects in lithium metal," Science, Vol. 356, Issue 6344, pp. 1254-1259 (2017).

[21]R. Shultz, "Lithium: Measurement of Young's Modulus and Yield Strength," –Fermilab-TM-2191, (2002).

[22]Robert B. Ross, <u>Metallic Materials Specification Handbook</u>, 4th ed; p. 15, (Chapman & Hall 1992).

[23]J. C. Boettger and S. B. Trickey, "Equation-of-State and Properties of Lithium," Physical Review B, Vol. 32, No. 6, 3391(1985).

[24] K. V. Khishchenko, "Equation-of-State for Lithium in Shock Waves," Mathematica Montisnigri, Vol XLI (2018).

[25] R. J. Naumann, "High Temperature Equation-of-State for Aluminum," NASA TN D-5892, Washington DC., (1972).

[26] P. Vinet, J. R. Smith, J. Ferrante, J. H. Rose, "Temperature effects on the universal equation-of-state of solids," Physical Review B, Vol. 35, No. 4, pp. 1945 (1987)

[27] Mofreh R. Zaghloul, "Dissociation and ionization equilibria of deuterium fluid over a wide range of temperatures and densities," Phys. Plasmas, 22, 062701 (2015).

[28] K. P. Huber and G. Herzberg, Molecular Spectra and Molecular Structure, IV. Constants of Diatomic Molecules (Litton.Educational Publishing, Inc., 1979).

[29] Mofreh R. Zaghloul, "On the ionization equilibrium of hot hydrogen plasma and thermodynamic-consistency of formulating finite internal partition functions," Phys. Plasmas, 17, 062701 (2010).

[30] Mofreh R. Zaghloul, ""Response to "Comment on "On the ionization equilibrium of hot hydrogen plasma and thermodynamic consistency of formulating finite partition functions," Phys. Plasmas, Vol. 17, 124706-3 (2010). DOI: 10.1063/1.3531707.

[31] Mofreh R. Zaghloul, "Inconsistency in Fermi's probability of the quantum states," Eur. Phys. J. H., 401 (2011). See also: M. R. Zaghloul, Eur. Phys. J. H., 38, 279 (2013).

[32] Mofreh R. Zaghloul, "Fundamental View on the Calculation of Internal Partition Functions Using Occupational Probabilities," Phys. Lett. A, 377, No. 15, 1119 (2013).

[33] David Salzmann, Atomic Physics in Hot Plasmas (Oxford University Press, New York, Oxford 1998).

[34] Robert E. Kelly, "Thermodynamics of blackbody radiation," Am. J. Phys. 49(8), 714-719

(1981).

[35]W. Ebeling, A. Förster, V. Fortov, V. Gryaznov, A. Polishchuk, <u>Thermophysical Properties of Hot Dense Plasmas</u>, (Leipzig Stuttgart 1991).

[36]T. Kahlbaum and A. Forster, "Generalized thermodynamic functions for electrons in a mixture of hard spheres: Application to partially ionized nonideal plasmas," Fluid Phase Equilibria, 76, 71 (1992).

[37]Mofreh R. Zaghloul, "Critical Parameters, Thermodynamic Functions and Shock Hugoniot of Aluminum Fluid at High Energy Density High Energy Density," High Energy Density Physics, Vol. 26, pp 8-15 (2018).

[38]L. Pauling, The Nature of the Chemical Bond (2nd ed.). Cornell University Press. LCCN 42034474 (1945).

[39]Hossein Eslami "A perturbed hard-sphere-chain equation-of-state for liquid metals," journal of Nuclear Materials," 336, 135-144 (2005).

[40]D. W. Jeppson, J. L. Ballif, W. W. Yuan, B. E. Chou, "Lithium literature review: Lithium's properties and interactions," HEDL-TME 78-15 uc-20 (1978).